\documentclass[letterpaper, 10 pt, conference]{ieeeconf}  

\IEEEoverridecommandlockouts                              

\let\oldTitle\title\renewcommand{\title}[1]{\oldTitle{\LARGE \bf #1}}
\newcommand{\mail}[1]{\tt\small #1}

    \usepackage[utf8]{inputenc}
    \usepackage[english]{babel}
    \usepackage[T1]{fontenc}
    \usepackage{amsfonts}
    \usepackage{mathtools}
    \usepackage{amssymb}
    \usepackage{stmaryrd}
    \usepackage{gensymb}
    \usepackage{hyperref}
    \usepackage{color}
    \usepackage{tikz}
    \usepackage{pgfplots}
    \usepackage{diagbox}
    \usepackage{dsfont}

    \newcommand{\iid}{\emph{i.i.d.}}
    \newcommand{\etal}{\emph{et. al}}
    \newcommand{\ie}{\emph{i.e}}
    \newcommand{\eg}{\emph{eg.}}
    \newcommand{\gft}{GFT}
    \newcommand{\eeg}{EEG}
    \newcommand{\rmse}{RMSE}
    \newcommand{\phd}{Ph.D.}
    
    \newcommand{\eqif}{\text{if}}
    \newcommand{\eqotherwise}{\text{otherwise}}
    
    \newcommand{\figref}[1]{Figure~\ref{#1}}
    
    \newcommand{\secref}[1]{Section~\ref{#1}}

    \newcommand{\pair}[2]{(#1; #2)}
    \newcommand{\intInterval}[2]{\{#1; \ldots; #2\}}
    \newcommand{\interval}[2]{[#1; #2]}
    \newcommand{\set}[1]{\{#1\}}
    \renewcommand{\~}[1]{\text{$\widetilde{#1}$}}
    \newcommand{\matij}[3]{\text{$#1_{#2; #3}$}}
    \newcommand{\veci}[2]{\text{$#1_{#2}$}}
    
    \newcommand{\G}{\text{$\mathcal{G}$}}
    \newcommand{\V}{\text{$\mathcal{V}$}}
    \newcommand{\E}{\text{$\mathcal{E}$}}
    \newcommand{\N}{\text{$N$}}
    \newcommand{\M}{\text{$M$}}
    \newcommand{\K}{\text{$K$}}
    
    \renewcommand{\P}{\text{$P$}}
    \renewcommand{\u}{\text{$u$}}
    \renewcommand{\v}{\text{$v$}}
    \newcommand{\w}{\text{$w$}}
    \newcommand{\I}{\textbf{I}}
    \newcommand{\W}{\textbf{W}}
    \newcommand{\D}{\textbf{D}}

    \newcommand{\T}{\textbf{T}}
    \newcommand{\NL}{\textbf{\L}}
    \newcommand{\x}{\textbf{x}}
    \newcommand{\s}{\textbf{s}}
    \newcommand{\X}{\textbf{X}}
    \newcommand{\Y}{\textbf{Y}}
    \newcommand{\covY}{\text{$\boldsymbol{\Sigma}_\Y$}}
    \newcommand{\eigval}[1]{\text{$\boldsymbol{\Lambda}_{#1}$}}
    \newcommand{\eigvec}[1]{\text{$\boldsymbol{\mathcal{X}}_{#1}$}}
    \newcommand{\el}{\text{$\lambda$}}
    \newcommand{\ev}{\text{$\boldsymbol{\chi}$}}
    
    \newtheorem{definition}{Definition}
    \newtheorem{remark}{Remark}
    
    \DeclareMathOperator*{\diag}{\mathrm{diag}}
    \DeclareMathOperator*{\triu}{\mathrm{triu}}
    \DeclareMathOperator*{\sign}{\mathrm{sign}}

    \newcommand{\setR}{\text{$\mathbb{R}$}}
    \newcommand{\vecZero}[1]{\text{$\mathds{O}_{#1}$}}
    
    \newcommand{\esp}[1]{\text{$\mathbb{E}[#1]$}}
    \newcommand{\tr}[1]{\text{$#1^\top$}}
    \newcommand{\distN}[2]{\text{$\mathcal{N}(#1, #2)$}}
    
    \title
    {
        Graph reconstruction from the observation of diffused signals*
        \thanks
        {
            *This work was supported by the European Research Council under the European Union's Seventh Framework Programme (FP7/2007-2013) / ERC grant agreement n\degree~290901.
            Additional thanks go to the CDI of Université Européenne de Bretagne and to the ISIS research group for the financial support.
        }
    }

    \author
    {
        Bastien Pasdeloup$^{1}$, Michael Rabbat$^{2}$, Vincent Gripon$^{1}$, Dominique Pastor$^{1}$ and Grégoire Mercier$^{1}$
        \thanks
        {
            $^{1}$Telecom Bretagne, UMR CNRS Lab-STICC, 655 Avenue du Technopole, 29280 Plouzané, France. \mail{name.surname@telecom-bretagne.eu}
        }
        \thanks
        {
            $^{2}$McGill University, ECE dept., 845 Rue Sherbrooke O, Montréal, Québec, Canada, QC H3A 0G4. \mail{name.surname@mcgill.ca}
        }
    }

\begin{document}
    \maketitle
    \thispagestyle{empty}
    \pagestyle{empty}
    \begin{abstract}
        \parskip=-\baselineskip
        \setlength\parindent{1.79cm}

    Signal processing on graphs has received a lot of attention in the recent years.
    A lot of techniques have arised, inspired by classical signal processing ones, to allow studying signals on any kind of graph.
    A common aspect of these technique is that they require a graph correctly modeling the studied support to explain the signals that are observed on it.
    However, in many cases, such a graph is unavailable or has no real physical existence.
    An example of this latter case is a set of sensors randomly thrown in a field which obviously observe related information.
    To study such signals, there is no intuitive choice for a support graph.
    In this document, we address the problem of inferring a graph structure from the observation of signals, under the assumption that they were issued of the diffusion of initially \iid{} signals.
    To validate our approach, we design an experimental protocol, in which we diffuse signals on a known graph.
    Then, we forget the graph, and show that we are able to retrieve it very precisely from the only knowledge of the diffused signals.
    
    \end{abstract}

    \section{INTRODUCTION}
        
        Signal processing on graphs has recently gained a lot of interest, and recent years have seen the appearance of numerous new techniques based on its principles.
        Although it was orignially seen as an extension of classical signal processing to more complex domains -- thus generalizing the Fourier approach --, applications now emerge in various fields such as wavelets on graphs, that find application in fields such as brain imagery \cite{Hammond2011}; or filtering, applied to big data analysis \cite{Sandryhaila2014}.
        Other examples are the recent work of Dong \etal{} \cite{Dong2014} that have studied the political orientations of the Swiss cantons from voting data; or the application of Abry \etal{} \cite{Abry2015} to photographic prints classification.
        
        One common aspect of all the existing applications is that they rely on the Graph Fourier Transform (\gft) of the graph, which is a cornerstone of signal processing on graphs.
        This particular transform is performed using the eigenvectors of a particular matrix that is dependent on the graph used to represent the support of the signals.
        However, such a matrix is not always available.
        In some even more complex -- and frequent -- cases, the graph has no real physical existence.
        Example of this are when sensors are thrown in a field, or are used to measure the brain activity (\eeg).
        There is an obvious link between the signal components observed in the various signals, but there is no intuitive notion of graph.
        
        To circumvent this problem, methods have recently been developed to allow one to recover a graph structure from the observation of signals.
        The idea behind this is to provide a structure that would \emph{naturally} represent the relationships between the signal components.
        Even in the case when the graph does not exist, such techniques should be able to provide a graph that captures these relationships, and would provide a support for the application of graph signal processing tools.
        
        One possible approach was proposed in the \phd{} dissertation of Costard \cite{Costard2014}.
        Her work is based on previous work from Dempster \cite{Dempster1972}, that estimates the covariance matrix of signals with the constraint that some values of its inverse --- called the precision matrix --- are null.
        This latter work has then allowed Wermuth \cite{Wermuth1976} to show that the non-null entries of this precision matrix correspond to edges of a graph representing the studied signals.
        Based on this result and on the additional works in \cite{Meinshausen2006, Banerjee2008} on model selection, Costard \cite{Costard2014} reviews existing methods for recovering a sparse precision matrix.
        Additionally, she introduces a new method called ABIGlasso, based on Bayesian inference techniques initialized using the Graphical lasso \cite{Friedman2008}.
        Related practical algorithms for estimating a sparse inverse covariance matrix were developed since then (\eg{} \cite{Hsieh2014}).
        
        Another existing approach to retrieve a graph from the observation of signals is the one introduced in the recent work of Dong \etal{} \cite{Dong2014}.
        The notion of smoothness of a signal on a graph was originally defined in \cite{Shuman2013} as a quantity representing the total variation of a signal with respect to adjacent nodes.
        Based on the assumption that adjacent nodes in a graph should share similar signal values --- \ie{} signals should be smooth --- Dong \etal{} propose a graph reconstruction technique that enforces this smoothness constraint.
        
        To the best of our knowledge, these two approaches are the main efforts that have been made to answer the problem of graph reconstruction from the observation of signals.
        However, for the approach of Costard, the author shows that her work is only applicable to a small number of variables \cite{Costard2014}.
        Concerning the approach of Dong \etal, they use the assumption that signals should be smooth on the recovered graph.
        In this paper, we propose a new method to recover a graph from signals, with different assumptions on the observed signals; that is to say they are issued from the diffusion of initially \iid{} signals in the graph.
        Our method consists of serching a solution in a system of linear constraints, which is of polynomial complexity \cite{Boyd2004} and can be done efficiently via the interior point method \cite{Karmarkar1984}.
        
        In this article, we first introduce in \secref{definitionsProperties} some definitions and properties that are needed for a complete understanding of our method.
        Then, in \secref{problem}, we introduce the setup that enables us to validate our approach, presented in \secref{graphReconstruction}.
        Finally, experimental evaluation is provided in \secref{evaluation}, and we discuss our method and future work in \secref{conclusions}.
        
    %


    \section{DEFINITIONS AND PROPERTIES}
    \label{definitionsProperties}
        
        In the context of signal processing on graphs, a graph is a mathematical structure used to represent the support of the observed signals.
        Particular instances of graphs can be used to represent the classical domains.
        For example, a chain graph can model the support of time-defined signals, and a grid can model the supporting pixels of an image.
        
        \begin{definition}[Graph]
            A graph \G{} is a pair \pair{\V}{\E} in which $\V = \intInterval{1}{\N}$ is a set of \N{} nodes and $\E = \V \times \V$ is a set of edges.
        \end{definition}
        
        With this definition, a signal on a graph can be seen as a value attached to every node in the graph:
        \begin{definition}[Signal]
            A signal \x{} on a graph \G{} of \N{} nodes is a vector in $\setR^\N$.
        \end{definition}
        
        A convenient way to represent a graph is through its adjacency matrix \W, defined as follows:
        \begin{definition}[Adjacency matrix]
            \begin{equation}
                \forall \u, \v \in \V : \matij{\W}{\u}{\v} \triangleq \left\{
                                                                          \begin{array}{cl}
                                                                              1 & \eqif \pair{\u}{\v} \in \E \\
                                                                              0 & \eqotherwise
                                                                          \end{array}
                                                                      \right.
                \;.
            \end{equation}
        \end{definition}
        
        In this article, we consider graphs with the following properties:
        \begin{itemize}
            \item Simple: $\forall \u \in \V : \matij{\W}{\u}{\u} = 0$.
            \item Undirected: $\forall \u, \v \in \V : \matij{\W}{\u}{\v} = \matij{\W}{\v}{\u}$.
            \item Connected: $\forall \u, \v \in \V : \exists \w_1 \ldots \w_k \in \V$ such that $\pair{\u}{\w_1} \in \E, \pair{\w_k}{\v} \in \E$ and $\forall i \in \intInterval{1}{k-1} : \pair{\w_i}{\w_{i+1}} \in \E$.
        \end{itemize}
        
        From the definition of the \W, we can define the diagonal degrees matrix \D{} of the graph, representing the degree of every node of \V:
        \begin{definition}[Degrees matrix]
            \begin{equation}
                \forall \u, \v \in \V : \matij{\D}{\u}{\v} \triangleq \left\{
                                                                          \begin{array}{cl}
                                                                              \sum\limits_{\w \in \V} \matij{\W}{\u}{\v} & \eqif \u = \v \\
                                                                              0                                          & \eqotherwise
                                                                          \end{array}
                                                                      \right.
                \;.
            \end{equation}
        \end{definition}
        
        Using the definitions of \W{} and \D{}, we can finally define the normalized Laplacian \NL{} of the graph.
        This is an operator analogous to the classical Laplacian operator \cite{Shuman2013}, that models the derivative of one diffusion in the graph:
        \begin{definition}[Normalized Laplacian]
            \begin{equation}
                \NL \triangleq \I_\N - \T
            \end{equation}
            with
            \begin{itemize}
                \item $\I_\N$ is the identity matrix of size \N.
                \item $\T \triangleq \D^{-\frac{1}{2}} \W \D^{-\frac{1}{2}}$ is called the diffusion matrix.
            \end{itemize}
        \end{definition}
        
        By construction, and with our assumptions on \G{}, \NL{} is a real symmetric matrix, and can thus be diagonalized into a basis of eigenvalues $\eigval{\NL} = \diag(\el_1 \ldots \el_\N)$ and associated eigenvectors $\eigvec{\NL} = \set{\ev_1; \ldots; \ev_\N}$.
        From \cite{Chung1997}, we know that the eigenvalues $\el_1 \leq \el_2 \leq \ldots \leq \el_\N$ of \NL{} have some interesting properties.
        One of them is that they are all in \interval{0}{2}, with $\el_\N = 2$ for and only for bipartite graphs (\ie{} graphs for which there exist two sets $\V_1$ and $\V_2$ such that $\V_1 \sqcup \V_2 = \V$, and such that $\forall \pair{\u}{\v} \in \E : (\u \in \V_1 \wedge \v \in \V_2) \vee (\v \in \V_1 \wedge \u \in \V_2)$).
        As a consequence, the eigenvalues of \T{} are in \interval{-1}{1}.
        
    %
    

    \section{PROBLEM SETUP}
    \label{problem}
        
        In this paper, we propose a method to recover a graph \G{} from the observation of signals diffused on it, with the assumption that they were initially \iid{}
        Hereafter, we denote as \X{} the $\N \times \M$ matrix of \iid{} signals, and $\Y = \T^\K \X$ the matrix of diffused signals, for a given value of \K.
        Thus, our problem becomes as follows:
        
        \emph{From the only observation of \Y, and under the assumption that \Y{} is issued from the diffusion of initially \iid{} signals, is it possible to recover \T, and thus \G?}
        
        More specifically, to validate our algorithm in \secref{graphReconstruction}, we create an experimental protocol consisting of five steps:
        \begin{enumerate}
            \item We randomly generate a graph \G, and compute the associated diffusion matrix \T.
                  In real applications, this graph is unknown, and is what we want to learn using our algorithm.
            \item We randomly generate \M{} \iid{} signals.
            \item For a chosen value of \K{}, we compute $\Y = \T^\K \X$.
            \item We use the algorithm presented in \secref{graphReconstruction} to recover an estimate \~\T{} for \T{} from only knowledge of \Y.
            \item We recover \~\G{} from \~\T.
        \end{enumerate}
        
        To study the reconstruction error, we use the root mean square error (\rmse) between the real and recovered adjacency matrices:
        \begin{definition}[Root mean square error]
            \begin{equation}
                \rmse(\W, \~\W) \triangleq \sqrt{\frac{1}{\N^2} \sum\limits_{i = 1}^\N \sum\limits_{j = 1}^\N | \matij{\W}{i}{j} - \matij{\~\W}{i}{j} |^2}
                \;.
            \end{equation}
        \end{definition}
        
        In this work, we only consider graphs such that the eigenvalues of \T{} have pairwise distinct absolute values.
        Note that this restricts the class of considered graphs to non-bipartite graphs that have one eigenvalue equal to $1$ and one equal to $-1$.
        
    %
    

    \section{GRAPH RECONSTRUCTION PROCESS}
    \label{graphReconstruction}
        
        In this section we detail our algorithm to recover an estimate \~\G{} for a graph \G, from only knowledge of \Y, as described in \secref{problem}.
        
        \subsection{Covariance of the diffused signals}
            
            Let \covY{} be the covariance matrix of \Y.
            With the assumption that the signals \Y{} are issued from the diffusion of initially \iid{} signals, we can make the following development:
            \begin{equation}
                \begin{array}{lll}
                    \covY & = & \esp{\Y \tr\Y} \\
                          & = & \esp{\T^\K \X \tr\X \tr{(\T^\K)}} \\
                          & = & \T^\K \esp{\X \tr\X} \tr{(\T^\K)} \\
                          & = & \T^{2\K} \;,
                \end{array}
            \end{equation}
            with the last simplification coming from --- with the \iid{} assumption --- $\esp{\X \tr\X} = \I_\N$ and from the observation that \T{} is symmetric.
            
            If we denote by \~\covY{} the empirical covariance of \Y{}, it is well known that \~\covY{} converges to \covY{} as \M{} tends to infinity.
            A consequence of this property is that we can obtain the eigenvectors of \~\T{} from the diagonalization of \~\covY{} --- since the eigenvectors of all powers of a matrix are identical --- provided that \M{} is big enough.
            
            However, this is not the case for the eigenvalues of \~\T. Since \~\covY{} is equal to $\~{\T^{2\K}}$, we have the property that the eigenvalues \eigval{\~{\T^{2\K}}} of $\~{\T^{2\K}}$ are all positive.
            Therefore, we can only obtain the absolute values of the eigenvalues of \~\T{} --- hereafter denoted \eigval{|\~\T|} --- through the 2\K-root of $\~{\T^{2\K}}$.
            
            In the remainder of this paper, we will assume that $\~\covY = \covY$.
            The impact of the noise when retrieving the covariance matrix will be briefly analyzed in \secref{evaluation}, and will be the main part of our future work.

        \subsection{Recuperation of the signs of \eigval{\~\T}}
        \label{optimizationProblem}
            
            At this point, we have a way to recover the eigenvectors \eigvec{\~\T} of \~\T{} and the absolute values \eigval{|\~\T|} of its eigenvalues, from only knowledge of the signals \Y{}, provided that \M{} is big enough.
            Now, we want to be able to recover a vector \s{} of signs ($-1$ or $+1$), such that $\eigval{\~\T} = \diag(\s) \eigval{|\~\T|}$ are eigenvalues such that $\~\T = \eigvec{\~\T} \eigval{\~\T} \tr{\eigvec{\~\T}}$ obeys some chosen properties:
            \begin{itemize}
                \item Values on the main diagonal of \~\T{} are all equal to 0.
                \item Off-diagonal values are either positive or null.
                \item \~\T{} is symmetric.
            \end{itemize}
            
            These three properties are by construction observed for any diffusion matrix in the class of graphs that we consider.
            Therefore, we will define a set of constraints that enforce these properties.
            
            The problem in retrieving \s{} directly is that the values we want \s{} to take are non-continuous, making the problem too difficult to solve.
            As a consequence, we will not put any constraint on the domain of \s, and will consider that it takes its values in $\setR^\N$.
            
            Based on the observation that the trace of \~\T{} should be equal to 0, and using the equality between the trace of a matrix and the sum of its eigenvalues, we create two sets of constraints:
            \begin{enumerate}
                \item \N{} equality constraints :
                      \begin{equation}
                          \forall i \in \intInterval{1}{\N} : \sum\limits_{j = 1}^\N \left( \matij{\~{\eigvec{\T}}}{i}{j} \right)^2 \veci{\s}{j} = 0
                          \;.
                          \label{equalityConstraints}
                      \end{equation}
                      These constraints enforce the main diagonal of \~\T{} to be null, when considering \s{} as its eigenvalues.
                \item $\frac{\N(\N-1)}{2}$ inequality constraints:
                      \begin{equation}
                          \forall \pair{i}{j} \in \triu(\N) : \sum\limits_{k = 1}^\N \matij{\~{\eigvec{\T}}}{i}{k} \matij{\~{\eigvec{\T}}}{k}{j} \veci{\s}{k} \geq 0
                          \;,
                          \label{inequalityConstraints}
                      \end{equation}
                      with $\triu(\N)$ being a function returning the indices of the triangular upper part of a $\N \times \N$ matrix.
                      These constraints enforce the off-diagonal elements to be either positive or null.
                      Moreover, considering only the triangular upper part enforces the symmetry of the recovered \~\T.
            \end{enumerate}
            
            Since a null vector of dimension \N{} is a trivial solution to the constraints, we use the observation that the eigenvalue associated to the constant-sign eigenvector should be equal to 1 \cite{Shuman2013}.
            Therefore, we add a final constraint to prevent this case:
            
            \begin{equation}
                \left[1; \vecZero{\N-1}\right]^\top \s = 1
                \;,
                \label{firstEigenvalueConstraint}
            \end{equation}
            with \vecZero{\N-1} being a null vector of size $\N-1$.
            
            By constraining the cells of \~\T{} through \eqref{equalityConstraints} and \eqref{inequalityConstraints} --- using its eigendecomposition $\~\T = \eigvec{\~\T} \diag(\s) \tr{\eigvec{\~\T}}$ --- we make \s{} converge toward a vector of values that should tend to the same sign as the expected eigenvalues.
            Finally, we only keep the signs of the eigenvalues, and inject them in the previously obtained \eigval{|\~\T|}:
            \begin{equation}
                \eigval{\~\T} = \sign(\s) \odot \eigval{|\~\T|}
                \;,
            \end{equation}
            with $\odot$ being the elementwise product of vectors, and $\sign(\s)$ being a function returning a vector in which the $i$-th component is equal to $-1$ when $\veci{\s}{i} < 0$, and $+1$ otherwise.
            
            We can note that, even though the problem is linear, we have a quadratic number of constraints in 2).
            We will show in \secref{evaluation} that we can limit ourselves to only a fraction of these constraints, thus reducing the complexity while still obtaining the same results.
            A solution to this set of linear constraints can be efficiently obtained using the interior point method \cite{Karmarkar1984}.
            
            \begin{remark}
                The restriction of our method to graphs having identical absolute eigenvalues, as described in \secref{problem}, can be understood here.
                As a matter of fact, if \T{} has eigenvalues of opposite signs, then when retrieving $\~\covY \sim \T^{2\K}$, these signs have disappeared.
                A consequence of this observation is that the diagonalization of \~\covY{} will not necessarily give the same eigenvectors as those of \T, since two distinct eigenspaces were merged into one.
            \end{remark}

        \subsection{Summary of the algorithm}
            
            In this section, we have proposed an algorithm to retrieve \~\T{} from the observation of diffused signals \Y, under the assumption that they were initially \iid{}
            Our method works in five steps, summarized as follows:
            \begin{enumerate}
                \item We compute \~\covY, the empirical covariance matrix of \Y.
                      For \M{} big enough, $\~\covY \sim \~{\T^{2\K}}$.
                \item The diagonalization of $\~{\T^{2\K}}$ gives a set of eigenvectors \eigvec{\~\T} and eigenvalues $\eigval{|\~{\T^{2\K}}|}$.
                \item The 2\K-root of the values in $\eigval{|\~{\T^{2\K}}|}$ are the absolute values of the eigenvalues of \~\T{} : \eigval{|\~\T|}.
                \item Find a solution to the constraints in \secref{optimizationProblem} to obtain \~\T.
                \item Since \T{} is a simple normalization by the degrees of \W{}, then it is null where and only where \W{} is null.
                      As a consequence, we fix a small $\varepsilon$ --- to circumvent machine precision limitations --- and retrieve \~\W{} from \~\T{} as follows:
                      \begin{equation}
                          \forall i, j \in \intInterval{1}{\N} : \matij{\~\W}{i}{j} = \left\{
                                                                                          \begin{array}{cl}
                                                                                              0 & \eqif \matij{\~\T}{i}{j} < \varepsilon \\
                                                                                              1 & \eqotherwise
                                                                                          \end{array}
                                                                                      \right.
                          \;.
                          \label{thresholding}
                      \end{equation}
            \end{enumerate}
            
            As we will show in \secref{evaluation}, this algorithm works well for the class of considered graphs.
            However, we make in this first work some assumptions on the number of observed signals, and on the distinguishability of the eigenvalues of the covariance matrix.
            Also, the value of \K{} is currently fixed in our work, and is common for every signal.
            We think that we can get rid of this additional parameter, and will study this hypothesis in future work.
            
        %
        
    %
    

    \section{EXPERIMENTAL EVALUATION}
    \label{evaluation}
        
        In this section, we first study the reconstruction performance --- and the associated execution time --- of our method for various graph orders, using the settings introduced in \secref{problem}.
        Then, we briefly study the impact of the value of \M{} --- and thus of the noise on the covariance matrix --- on the successfulness of our algorithm.
        
        To perform our tests, we make some experimentation choices, such as the use of an Erdős-Rényi model \cite{Erdos1959}, with edge probability $\P = 0.3$, for \G; and the choice of a normal distribution \distN{0}{1} for \X.
        Similar results are observable for different graph definitions --- such as random geometric graphs --- and choices of distribution.
        
        All tests in this document are performed using the CVX package \cite{Grant2014} with solver SDPT3 \cite{Toh1999} for MATLAB \cite{MATLAB2012}, with default parameters.
        
        \subsection{Reconstruction performance of the method}
            
            We have seen in \secref{optimizationProblem} that the number of inequality constraints describing the matrix to reconstruct was quadratic in the dimension of the signals.
            For this reason, we limit ourselves a randomly selected subset of them.
            By doing so, we slightly increase the number of possible solutions.
            However, it appears that this class of solution vectors has still a very low cardinality --- in fact it is most of the time unique --- and that it is sufficient to guide our solver by constraining only a small number number of off-diagonal cells to be positive.
            
            \figref{errorAsFunctionOfGraphOrder} shows the mean \rmse{} between the initial \W{} and the reconstructed estimate \~\W{}, for various ratios of kept constraints, \ie{} for various values of $\alpha$ in $\alpha\frac{\N(\N-1)}{2}$.
            
            Here, we use an oracle to find the best $\varepsilon$ possible in \eqref{thresholding}, although the gap with the real values of \~\T{} is generally high and it is quite easy to distinguish noise from data.
            We make this choice because we want to study the reconstruction performance of our method.
            In real applications, $\varepsilon$ should impose a degree of sparsity on the reconstructed matrix.
            
            \begin{figure}[h!]
                \centering
                \begin{tabular}{|l|c|c|c|c|c|c|}
                    \hline
                    \diagbox{\N}{$\alpha$} & 0 & $\frac{1}{2\N}$ & $\frac{1}{\N}$ & $\frac{1}{4}$ & $\frac{1}{2}$ & 1 \\
                    \hline
                    25 & 0.3835 & 0.052153 & 0 & 0 & 0 & 0 \\
                    \hline
                    50 & 0.29981 & 0 & 0 & 0 & 0 & 0 \\
                    \hline
                    100 & 0.2739 & 0 & 0 & 0 & 0 & 0 \\
                    \hline
                \end{tabular}
                \caption
                {
                    \rmse{} as a function of the graph order, for various ratios of kept constraints.
                    Mean for 20 randomly generated graphs.
                }
                \label{errorAsFunctionOfGraphOrder}
            \end{figure}
            
            We can see from this table that our algorithm works quite well, and succeeds in reconstructing the original graph in 100\% cases in our tests when we keep a number of inequality constraints over $\frac{1}{\N}$.
            We can also see that even with a smaller number of constraints, the method is still robust, but some errors start to appear, mainly for smaller graphs.
            However, the first column shows that keeping some of these inequality constraints are essential to guide the solver toward the correct solution.
            
\begin{figure}[h!]
                \centering
                \scalebox{0.54}{\Large{\input{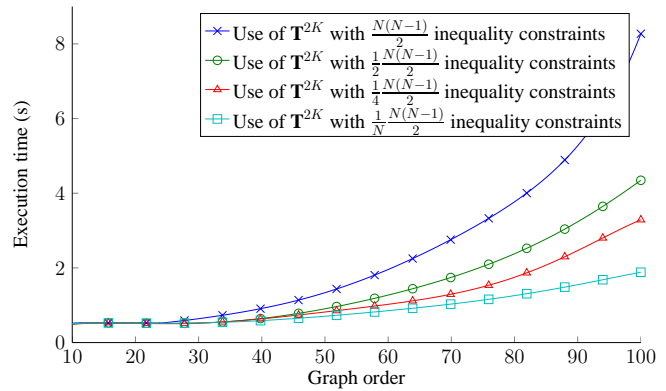}}}
                \caption
                {
                    Execution time as a function of the graph order, for various ratios of kept constraints.
                    Mean for 20 randomly generated graphs.
                }
                \label{timeAsFunctionOfGraphOrder}
            \end{figure}

            Because removing some constraints does not overly increase the error, we study the impact it has on the execution time of our algorithm.
            \figref{timeAsFunctionOfGraphOrder} depicts the time required by our algorithm to reconstruct the graph, as a function of the graph order, for the same ratios of kept constraints.
            
            As expected, removing some inequality constraints in \eqref{inequalityConstraints} considerably reduces the execution time of our algorithm.
            As a matter of fact, we can remove a nonlinear number of constraints while keeping the same performance, thus making the number of inequality constraints linear.
            A consequence of this is that our method is now based on finding a solution to a linear number of linear constraints.
            Therefore, it can be applied to graph reconstruction with high dimension signals.

        \subsection{Impact of the noise on the reconstruction}
            
            At this point, we have made one strong assumption that the number \M{} of observed signals is big enough to consider $\~\covY = \covY$.
            Although it is true in the limit case of an infinite number of signals, in real applications there is some noise in the empirical covariance matrix.
            
\begin{figure}[h!]
                \centering
                \scalebox{0.54}{\Large{\input{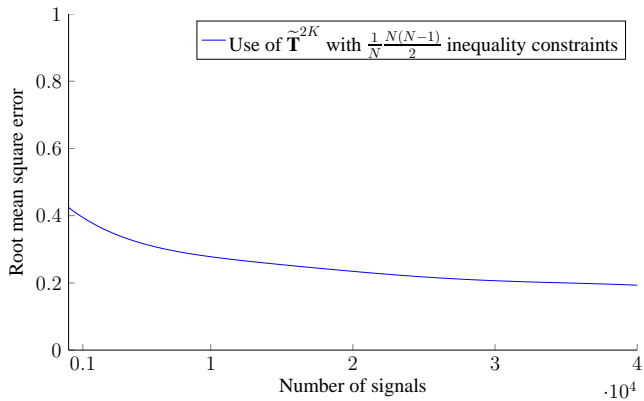}}}
                \caption
                {
                    \rmse{} as a function of the number of signals, for graphs of 15 nodes, when using \~\covY{} instead of \covY, with a ratio of inequality constaints $\alpha = \frac{1}{\N}$.
                    Mean for 1000 randomly generated graphs.
                }
                \label{noiseAsFunctionOfNumberOfSignals}
            \end{figure}

            \figref{noiseAsFunctionOfNumberOfSignals} shows the impact of the value of \M{} on the \rmse{}, for randomly generated graphs of 15 nodes.
            
            In addition to these curves, it is interesting to remark that the \rmse{} when performing our tests is generally either 0 or very close, or around $0.5$.
            Although the mean error decreases as the \M{} increases, the previous observation shows that noise when retrieving \~\covY{} is the main problem of our method, and will therefore be the main focus of our future work.
            
            One promising direction comes from the observation that most of the error is concentrated in the eigenvectors of \~\covY{} and not in its eigenvalues.
            Therefore, we might be able to use the fact that the first eigenvector of \~\T{} should be of a certain form --- from the observation that the components of the first eigenvector of \NL{} are values that are function of the graph degrees --- to realign the eigenvectors of \~\covY{} properly.
            
        %
        
    %
    

    \section{CONCLUSIONS}
    \label{conclusions}
        
        In this paper, we have introduced a scalable method to retrieve a graph from the observation of signals diffused on it.
        We have experimentally shown that we could reduce the number of inequality constraints in our problem, thus making our method as complex as finding a solution to a linear number of linear constraints.
        
        The main limitation of our work is currently the impact of the noise on the reconstruction of the covariance matrix of the observed signals.
        For this reason, this will be the core aspect of our future studies.
        Additionally, we want to get rid of the parameter \K, describing the number of diffusions of the observed signals, in our algorithm.
        This would enable us to consider signals observed at different points in time in an unified framework.


\begin{thebibliography}{99}

    \bibitem{Abry2015}
    
        P. Abry, S. G. Roux, H. Wendt, P. Messier, A. G. Klein, N. Tremblay, P. Borgnat, S. Jaffard, B. Vedel, J. Coddington, and L. A. Daffner,
        Multiscale Anisotropic Texture Analysis and Classification of Photographic Prints,
        IEEE Signal Processing Magazine,
        2015.

    %
    

    \bibitem{Banerjee2008}
        
        O. Banerjee, L. El Ghaoui and A. d'Aspremont,
        Model Selection Through Sparse Maximum Likelihood Estimation for Multivariate Gaussian or Binary Data,
        J. Mach. Learn. Res., vol. 9, pp. 485--516,
        2008
    
    %


    \bibitem{Boyd2004}
        
        S. Boyd and L. Vandenberghe,
        Convex Optimizationn
        Cambridge University Press,
        2004
    
    %


    \bibitem{Chung1997}
    
        F. R. K. Chung,
        Spectral Graph Theory,
        American Mathematical Society,
        1997.
        
    %


    \bibitem{Costard2014}
        
        A. Costard,
        Estimation de la structure d'indépendance conditionnelle d'un réseau de capteurs. Application à l'imagerie médicale.
        Ph.D. dissertation, Université de Grenoble and Ecole Normale Supérieure de Lyon,
        2014.
        
    %


    \bibitem{Dempster1972}
        
        A. P. Dempster,
        Covariance Selection,
        Biometrics,
        1972.
        
    %


    \bibitem{Dong2014}
        
        X. Dong, D. Thanou, P. Frossard and P. Vandergheynst,
        Learning Graphs from Signal Observations under Smoothness Prior,
        CoRR,
        2014.
        
    %
    

    \bibitem{Erdos1959}
        
        P. Erdős and A. Rényi,
        On Random Graphs I.,
        Publicationes Mathematicae,
        1959.
        
    %
    

    \bibitem{Friedman2008}
        
        J. Friedman, T. Hastie, and R. Tibshirani,
        Sparse inverse covariance estimation with the Graphical lasso,
        Biostatistics,
        2008.
        
    %


    \bibitem{Hsieh2014}
    
        C.-J. Hsieh, M. A. Sustik, I. S. Dhillon and P. Ravikumar,
        QUIC: Quadratic Approximation for Sparse Inverse Covariance Estimation,
        Journal of Machine Learning Research, vol. 15, pp. 2911-2947,
        2014
        
    %


    \bibitem{Grant2014}
        
        M. Grant and S. Boyd,
        CVX: Matlab Software for Disciplined Convex Programming, version 2.1,
        \url{http://cvxr.com/cvx},
        2014.
        
    %
    

    \bibitem{Hammond2011}
    
        D K. Hammond, P. Vandergheynst and R. Gribonval,
        Wavelets on Graphs via Spectral Graph Theory,
        Applied and Computational Harmonic Analysis, vol. 30, pp. 129--150,
        2011.
    
    %


    \bibitem{Karmarkar1984}
    
        N. Karmarkar,
        A new polynomial-time algorithm for linear programming,
        Combinatorica, vol. 4, pp. 373-395,
        1984
        
    %


    \bibitem{MATLAB2012}
    
        MATLAB,
        version 7.14.0 (R2012a),
        The MathWorks Inc., Natick, Massachusetts,
        2012
        
    %


    \bibitem{Meinshausen2006}
    
        N. Meinshausen and P. Bühlmann,
        High-dimensional graphs and variable selection with the Lasso,
        The Institute of Mathematical Statistics, The Annals of Statistics, vol. 34, pp. 1436--1462,
        2006
        
    %


    \bibitem{Sandryhaila2014}
        
        A. Sandryhaila and J. M. F. Moura,
        Big Data Analysis with Signal Processing on Graphs,
        IEEE Signal Processing Magazine, vol. 31, pp. 80--90,
        2014.
        
    %


    \bibitem{Shuman2013}
    
        D. I. Shuman, S. K. Narang, P. Frossard, A. Ortega and P. Vandergheynst,
        The Emerging Field of Signal Processing on Graphs: Extending High-Dimensional Data Analysis to Networks and Other Irregular Data Domains,
        CoRR,
        2013.
        
    %


    \bibitem{Toh1999}
        
        K. C. Toh, M. J. Todd and R. H. Tütüncü,
        SDPT3 -- a MATLAB software package for semidefinite programming,
        Optimization methods and software,
        1999
        
    %


    \bibitem{Wermuth1976}
        
        N. Wermuth,
        Analogies between multiplicative models in contingency tables and covariance selection,
        Biometrics,
        1976.
        
    %

    \end{thebibliography}
\end{document}